\documentclass[jkps,twoside,twocolumn,showpacs,showkeys] {revtex4}
\usepackage{graphicx}

\begin{document}
\title[]{Statistical Properties of the Returns of Stock Prices of International Markets}

\author{GabJin \surname{Oh}}
\email{gq478051@postech.ac.kr} \affiliation{Asia Pacific Center for Theoretical Physics, National Core Research Center for Systems Bio-Dynamics and Department of Physics, Pohang University of Science
and Technology, Pohang, Gyeongbuk 790-784}

\author{Seunghwan \surname{Kim}}
\email{swan@postech.ac.kr} \affiliation{Asia Pacific Center for Theoretical Physics, National Core Research Center for Systems Bio-Dynamics and Department of Physics, Pohang University of Science
and Technology, Pohang, Gyeongbuk 790-784}

\author{Cheol-Jun \surname{Um}} \email{shunter@cup.ac.kr} \affiliation{Department of Healthcare Management, Catholic University of Pusan, Busan 609-757}

\received{16 September 2005}

\begin{abstract}
We investigate statistical properties of daily international
market indices of seven countries, and high-frequency $S\&P500$
and KOSDAQ data, by using the detrended fluctuation method and the
surrogate test. We have found that the returns of international stock
market indices of seven countries follow a universal power-law
distribution
 with an exponent of $\zeta \approx 3$, while the Korean stock market follows an exponential distribution with an
 exponent of $\beta \approx 0.7$. The Hurst exponent analysis of the original return, and its magnitude and sign series, reveal
 that the long-term-memory property, which is absent in the returns and sign series,
 exists in the magnitude time series with $0.7 \leq H \leq 0.8$. The surrogate test shows that the magnitude time series
 reflects the non-linearity of the return series, which helps to reveal
 that the KOSDAQ index, one of the emerging markets, shows higher volatility than a mature market such as the {S\&P} 500 index.\\


\end{abstract}

\maketitle

\section{INTRODUCTION}
Up to now, numerous studies analyzing financial
time series have been carried out to understand the complex economic systems made up of
diverse and complicated agents  \cite{R1}. The statistical analysis of economic or
financial time series exhibits features different from
the random-walk model based on the efficient market hypothesis (EMH), which
are called stylized facts [2-15]. Previous studies
showed that the returns of both stocks and foreign exchange rate have
a variety of stylized facts. For example, the distribution of
financial time series follows a universal power-law distribution with an
exponent $\zeta \approx 3$ [3-7]. While the temporal correlation of returns
follows the process of random walks, the volatility of returns shows a
long-term-memory property [12-15]. However, recent work has revealed that the distribution function of returns in emerging markets
follows an exponential distribution, while the mature markets follow a power-law distribution with an exponent $\zeta \approx 3 $ \cite{R11}.

In this paper, we use Detrended Fluctuation Analysis (DFA),
which was introduced by Peng {\em et al.} to find the long-term-memory
property of time series data \cite{R16} and utilizes the surrogate
test method proposed by Theiler {\em et al.} to measure the
non-linearity of time series \cite{R17}. We study daily
international market indices of seven countries from 1991 to 2005, 
the high-frequency $S\&P$ 500 5-minute index from 1995 to
2004, and the high-frequency KOSDAQ 1-minute index from 1997 to
2004, to investigate diverse time characteristics of financial
market indices.

We found that the returns of international stock market indices of
seven countries follow a universal power-law distribution with an
exponent of $\zeta \approx 3$, while the Korean stock market
follows an exponential distribution with an exponent of $\beta
\approx 0.7$. For a more detailed statistical analysis, the original return
time series is divided into magnitude and sign time series, and
the corresponding Hurst exponents are computed. The Hurst exponent
analysis of the original return, and its magnitude and sign time series, 
reveal that the long-term-memory property, which is absent in the
return and sign time series, exists in the magnitude time series with
$0.7 \leq H \leq 0.8$.

In order to test the nonlinearity of the time series, the
surrogate test is performed for all time series. We find that the
magnitude time series reflects the non-linearity of the return
series, which helps to reveal that the KOSDAQ index, one of the emerging markets, shows higher volatility than a mature market such as the {S\&P} 500 index.

In the next section, we explain the market data used in our
investigations. In Section \ref{sec:METHODS}., we introduce the methods of the
surrogate test and detrended fluctuation analysis (DFA). In Section \ref{sec:RESULTS}., the results of the statistical analysis for various
time series of the market data are presented. Finally, we end with
a summary of our findings.

\section{DATA}
We use the return series in eight daily international market indices 
of seven countries from 1991 to 2005, the $S\&P$ 500 index (5 minutes) from 1995 to 2004, and the KOSDAQ index (1 minute) 
from 1997 to 2004. The seven countries are France (CAC40), Germany (DAX), United Kingdom (FTSE100), Hong Kong (HangSeng), KOREA (KOSPI), 
America (NASDAQ), Japan (Nikkei225), and America (S\&P 500). We make use of the normalized return often used 
in the financial time series analysis instead of the stock prices.
Let $y_1,y_2,....y_n$, 
be the daily stock prices. 
The normalized return $R_{t}$ at a given time t is defined by

\begin{eqnarray}\label{e1}
&& {r_t} = \ln{y_{t+1}}- \ln{y_{t}}, \nonumber \\
&& {R_t} \equiv  \frac{\ln{y_{t+1}}- \ln{y_{t}}}{\sigma(r_t)},\label{e1}
\end{eqnarray}
where $\sigma(r_t)$ is the standard deviation of the return. The normalized returns $R_t$
 are divided into magnitude and sign series by using the following relation:

\begin{equation}\label{e2}
{R_{k,t}}= |R_{k,t}|  \times Sign_{k,t}, \\
\end{equation}
where $R_{k,t}$ is the return series of the k-th market index calculated by the log-difference,
$ |R_{k,t}|$ the magnitude series of the returns of the k-th market index, and $Sign_{k,t}$ the sign series with $+1$ for the upturn and $-1$ for the downturn.
Note that the magnitude series $|R_t|$ from taking the absolute value of 
the return measures the size of the return change, and the sign series $Sign_t$ 
measures the direction of the change. The volatility of the returns
can be studied though the magnitude series $|R_t|$.

\section{METHODS}
\label{sec:METHODS}

\subsection{Surrogate Test}

The surrogate test method was first proposed by Theiler {\em et al.} to prove the non-linearity 
contained in the time series \cite{R17}. The surrogate data test can be explained by the following four steps \cite{R16}.
First, a null hypothesis is made and the features of the linear process following the hypothesis are defined. In general, the linearity uses the mean, the variance, and the autocorrelation of the original time series. 
The surrogate data are randomly generated but retain the autocorrelation function, the mean, and the variance of the original data.
In the second step, the surrogate data are created through the Fast Fourier Transform(FFT) method. 
Let $r_n$ be the original time series. The Fourier Transform $r_k$ of $r_n$ is given by

\begin{equation}\label{e3}
{r_k}= \frac{1}{N}\sum_{n=1}^{N} r_n e^{i2 \pi nk/N}.
\end{equation}
Then, $r_k$ is multiplied by random phases,

\begin{equation}\label{e4}
\tilde{r_k}= {r_k} e^{i \phi_k},
\end{equation}
where $\phi_k$ is uniformly distributed in [0, 2$\pi$]. The inverse FFT of $\tilde{r_k}$ gives the surrogate data 
retaining the linearity in the original time series, 

\begin{equation}\label{e5}
r_{n}^{'} = \frac{1}{N}\sum_{k=1}^{N} \tilde{r_k}e^{-i2\pi nk/N}.
\end{equation}
In the third step, non-linear measurements with the entropy, the dimension, and Lyapunov exponents 
are performed for the original data and the surrogate data, respectively. Finally, the difference 
in measurements of the original data and the surrogate data is tested for significance. If significant, the hypothesis will be rejected 
and the original data are regarded as having non-linearity.

\subsection{Detrended Fluctuation Analysis}

The typical methods to analyze the long-term-memory property in the time series data are 
largely classified into three types: the re-scaled range analysis (R/S) method 
proposed by Mandelbrot and Wallis \cite{R19}, the modified R/S analysis by Lo {\em et al.} \cite{R18}, 
and the DFA (detrended fluctuation analysis) method by Peng {\em et al.} \cite{R20}. In this paper, 
the DFA method is used due to its effectiveness even for the absence of long-term memory. The Hurst exponent can be calculated 
by the DFA method through the following process.
 
Step (1): The time series after the subtraction of the mean are accumulated as follows:

\begin{equation}\label{e6}
y(i)= \sum_{i=1}^{N} [x(i) - \bar{x}],
\end{equation}
where $x(i)$ are the i-th time series, and $\bar{x}$ is the mean of the whole time series. 
This accumulation process is one that changes the original data into a self-similar process.

Step (2): The accumulated time series are divided into boxes of the same length $n$. 
In each box of length $n$, the trend is estimated through the ordinary least square method, called DFA(m), where m is the order of fitting. 
In each box, the ordinary least square line is expressed as $y_{n}(i)$. By subtracting $y_{n}(i)$ 
from the accumulated $y(i)$ in each box, the trend is removed. This process is applied to every box 
and the fluctuation magnitude is calculated by using 

\begin{equation}\label{e7}
F(n)= \sqrt{\frac{1}{N} \sum_{i=1}^{N} [y(i) - y_{n}(k)]^2}.
\end{equation}
The process of Step (2) is repeated for every scale, 
from which we obtain a scaling relation
\begin{equation}\label{e8}
F(n) \approx  cn^{H},
\end{equation}
where H is the Hurst exponent. The Hurst exponent characterizes the correlation of time series with three different properties. 
If $0 \leq H < 0.5$, the time series is anti-persistent. If $0.5 < H \leq 1$, it is persistent. In the case of H = 0.5, 
the time series correspond to random walks.

\section{Results}
\label{sec:RESULTS}

\begin{figure}[tb]
\includegraphics[height=4cm, width=8cm]{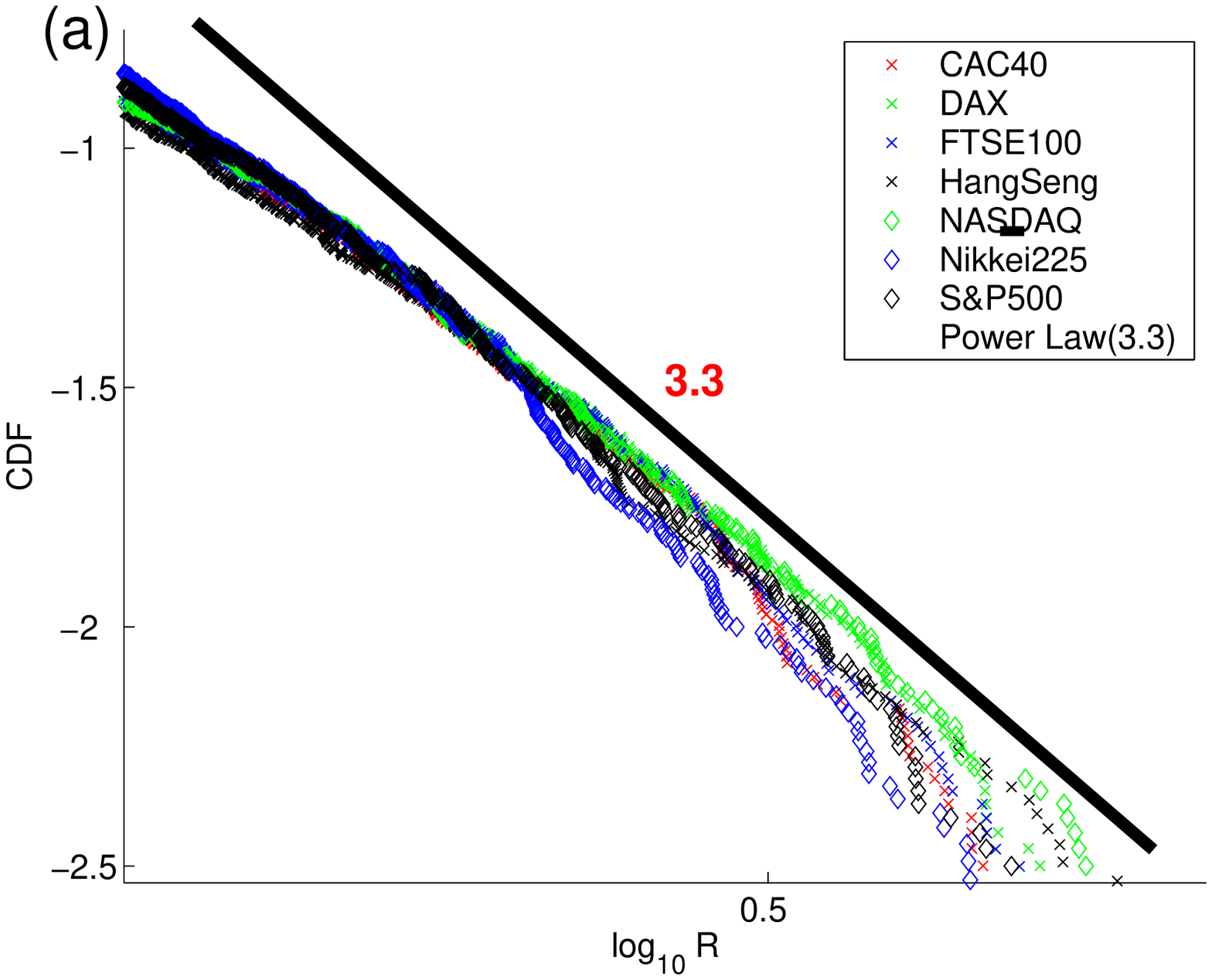}
\includegraphics[height=4cm, width=8cm]{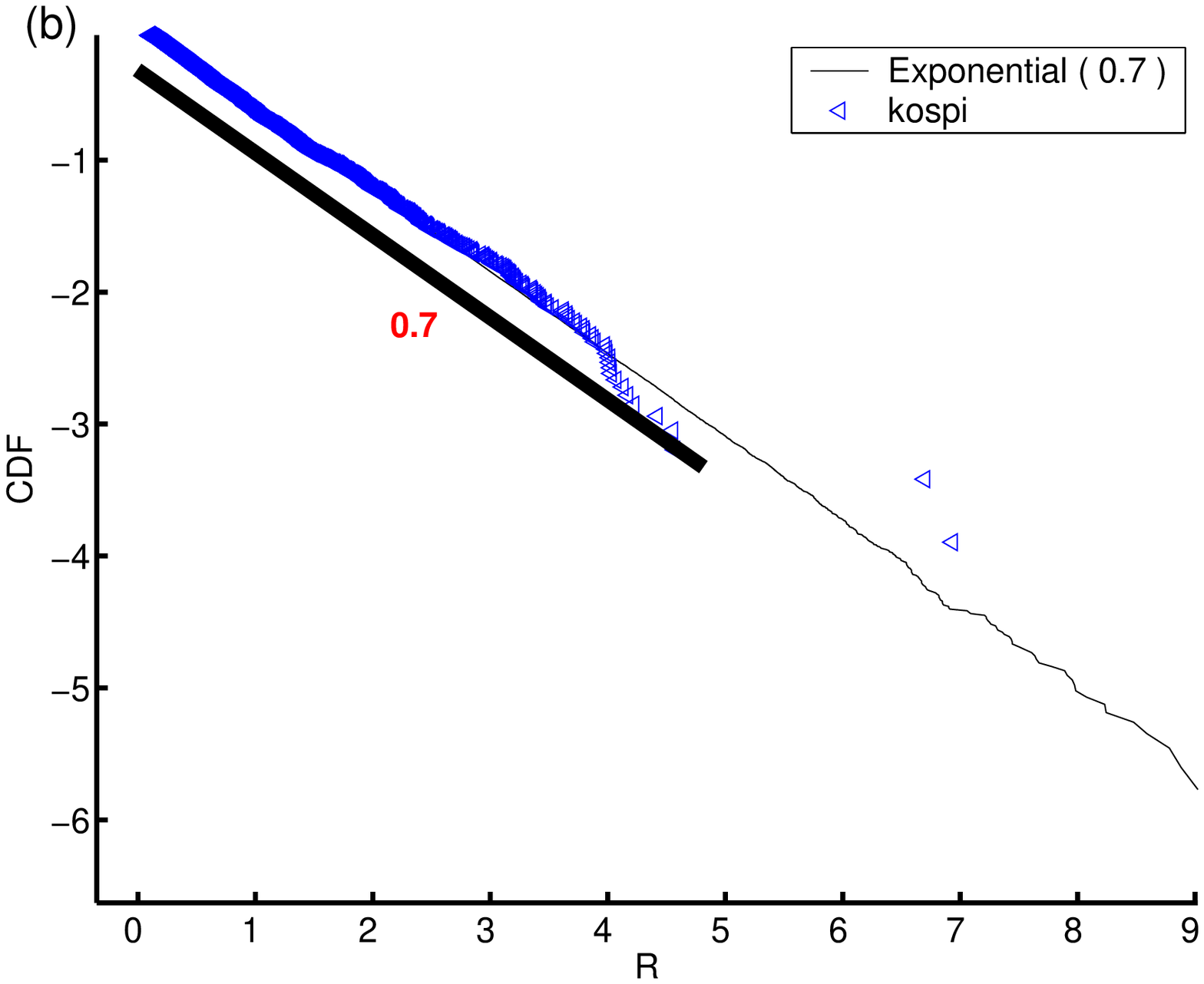}
\caption{Cumulative distribution function (CDF) $P(R_{t} > R)$ of normalized returns time series $R_{t}$. (a) Normalized return distribution of international 
market indices of six countries, excluding Korea, from January 1991 to May 2005 in a log-log plot. (b) Linear-log plot for the KOSPI 
index.}
\end{figure}

\begin{figure}[tb]
\includegraphics[height=5cm,width=8cm]{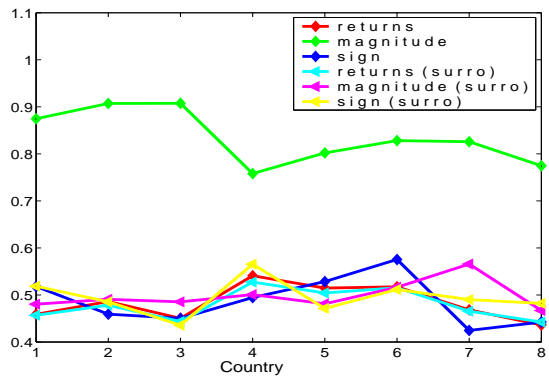}
\caption[0]{Hurst exponents of international market indices [1: France (CAC40), 2: Germany (DAX), 3: United Kindom (FTSE 100), 
4: Hong Kong (HangSeng), 5: Korea (KOSPI), 6: America (Nasdaq), 7: Japan (Nikkei 225), 8: America ($S\&P$ 500)] from the return, magnitude and sign time series. 
The notation (surro) denotes the corresponding surrogate data.}
\end{figure}

\begin{figure}[tb]
\includegraphics[height=7cm,width=8cm]{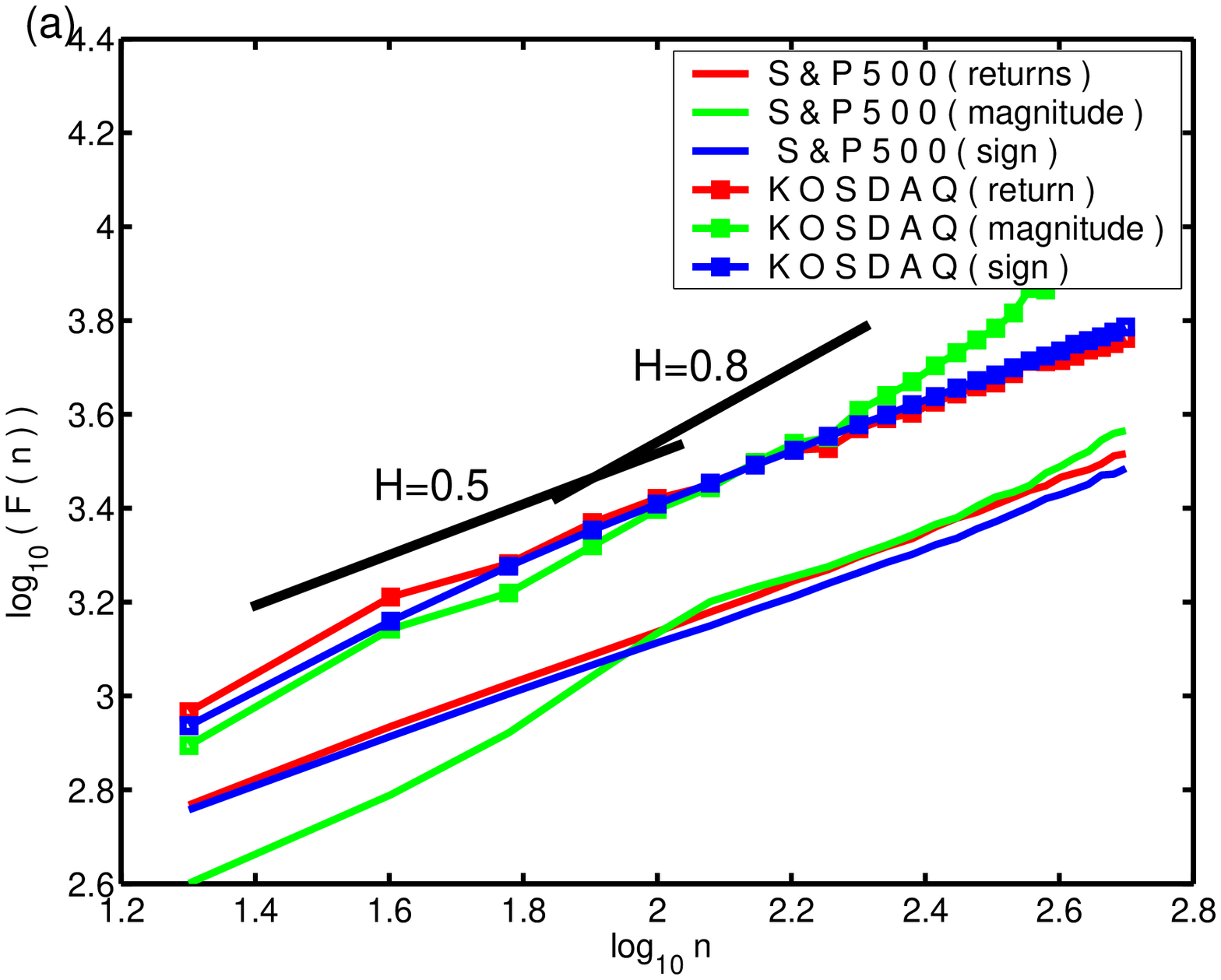}
\includegraphics[height=7cm,width=8cm]{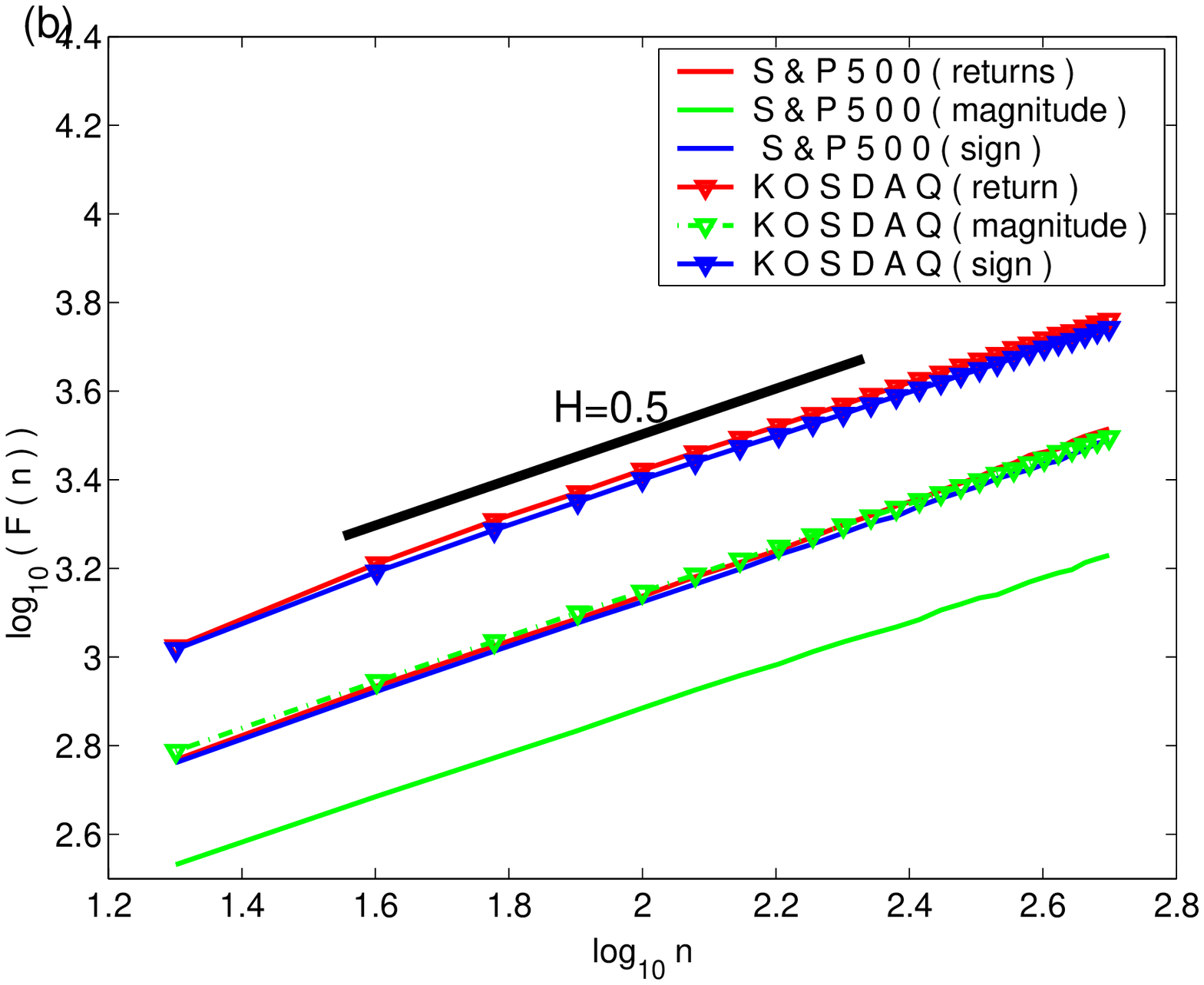}
\caption[0]{(a) Hurst exponent of the $S\&P$ 500 5-minute index and the KOSDAQ 1-minute index
with the time series of the returns divided into magnitude and sign time series. (b) Hurst exponent of the surrogate data of 
the $S\&P$ 500 and KOSDAQ indicies.}
\end{figure}

\begin{figure}[tb]
\includegraphics[height= 10cm, width=9cm]{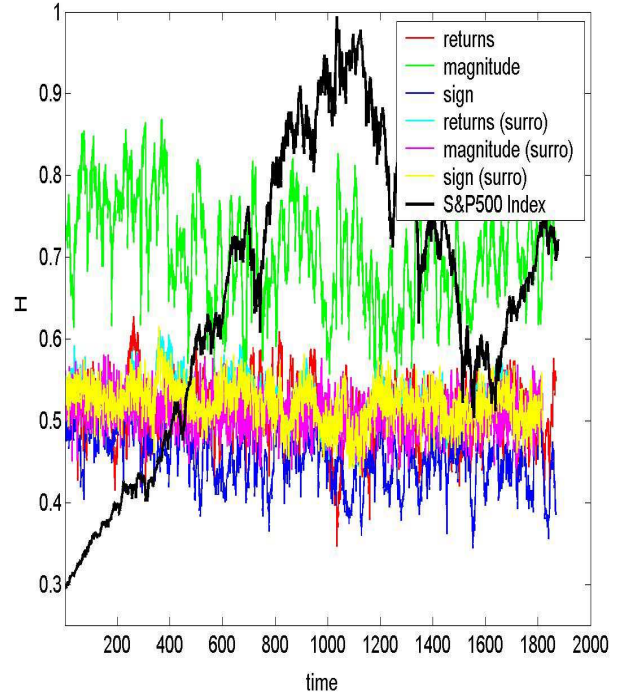}
\caption[0]{Hurst exponent of $S\&P$ 500 5-minute index returns divided into magnitude and sign: 
the black solid line denote the price of $S\&P$ 500 from 1995 to 2004. The other lines denotes the Hurst exponents corresponding to the returns, 
sign and magnitude time series and the Hurst exponents of the returns, sign and magnitude time series of the surrogate data. The
notation (surro) denotes the corresponding surrogate data.}
\end{figure}

\begin{figure}[tb]
\includegraphics[height= 10cm, width=9cm]{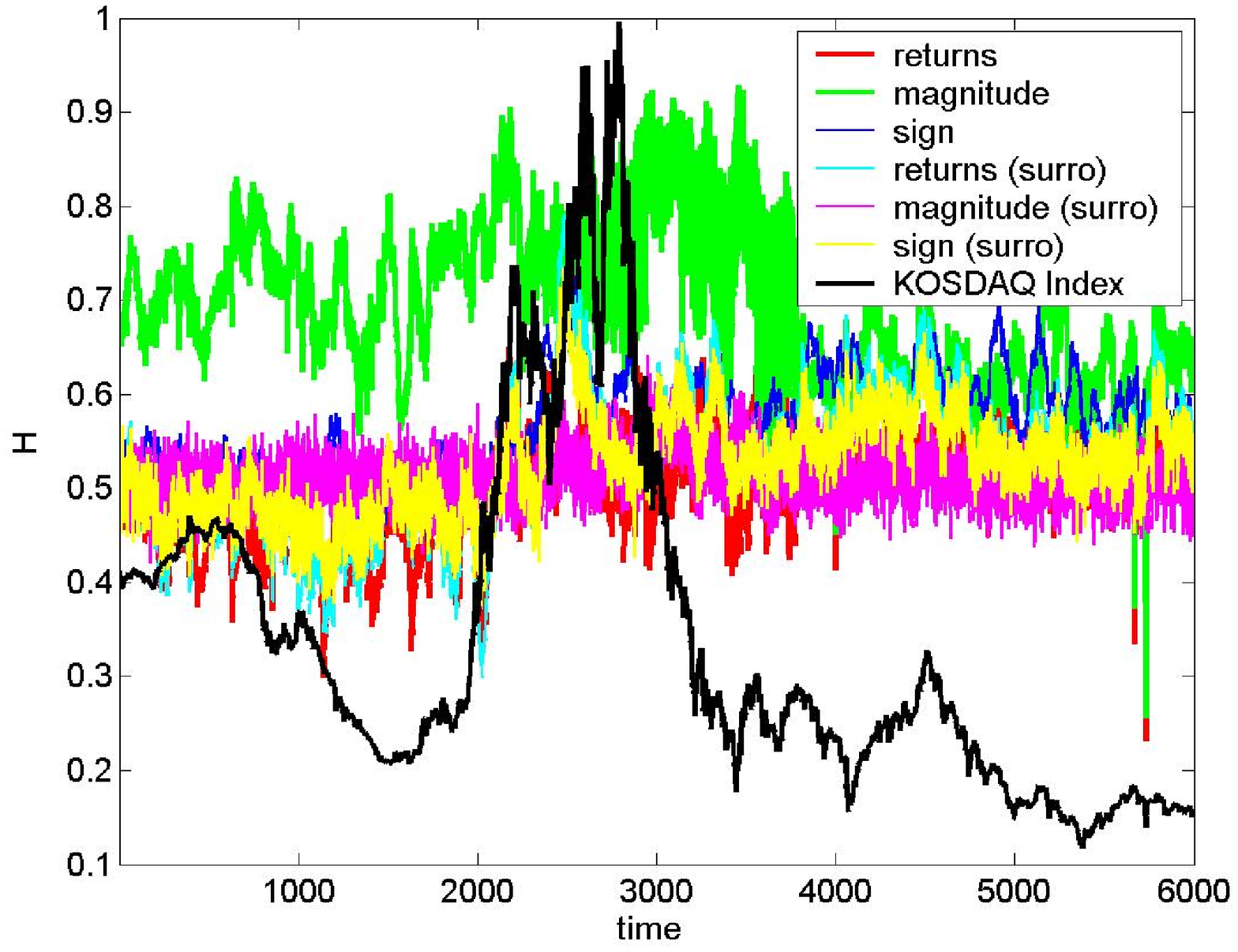}
\caption[0]{Hurst exponent of the KOSDAQ 1-minute index returns divided into magnitude and sign series. 
The solid black line shows the KOSDAQ index from 1997 to 2004. The other lines denote the Hurst exponents
for the returns, sign and magnitude time series and the corresponding surrogate data. The
notation (surro) denotes the corresponding surrogate data.}
\end{figure}

In this section, we analyze the statistical features of daily international market indices of seven countries 
from January 1993 to May 31, 2005, the $S\&P$ 500 5-minute index from 1995 to 2004, and 
the KOSDAQ 1-minute index from 1997 to 2004. 
We present the results of the statistical features such as the cumulated distribution function (CDF) and the time correlation of the various 
financial time series. Figure 1(a) is a log-log plot of the cumulative distribution function 
of the market indices of six countries, excluding Korea, from January 1991 to May 2005. Figure 1(b) is a linear-log plot of the distribution function 
of the KOSPI index.

In Figure 1(a), we find that the tail exponents of the indices of all the countries except Korea follow a universal power-law distribution 
with an exponent $\zeta \approx 3$. However, in Figure 1(b), we find that the Korean stock market follows an exponential distribution with $\beta \approx 0.7$, 
not a power-law distribution. These results indicate that the distribution of returns in the KOSPI index, that belongs to the emerging markets, does not follow a power-law distribution with 
the exponent $\zeta \approx 3$.

Figure 2 shows the Hurst exponents for the returns of each international market index, calculated from the return, magnitude and sign time series. The long-term-memory property is not found for the return and sign series with $H \approx 0.5$. 
However, we find that the magnitude time series has a long-term-memory property with $H \approx 0.8$. 
The surrogate test plots denoted as (surro) in Figure 2 show that the magnitude time series reflects the non-linearity of the original returns,
while the sign time series shows the linearity of the original returns.

In order to investigate the scaling in high-frequency data, 
we use the $S\&P$ 500 5-minute index from 1995 to 2004 and the KOSDAQ 1-minute index from 1997 to 2004. 
Figure 3(a) shows the Hurst exponents of the return, magnitude and sign series for the $S\&P$ 500 5-minute index and the KOSDAQ 
1-minute index. As for international stock market indices,
the sign series corresponds to random walks ($H \approx 0.5$), but the magnitude series
has a long-term-memory property ($0.7 \leq H \leq 0.8$). Figure 3(b) shows that all Hurst exponents of the corresponding surrogate data follow random walks.

In order to find the time evolution of the Hurst exponent, we also investigated the time series by shifting the $S\&P$ 500 5-minute index
and KOSDAQ 1-minute index by 500 minutes and 100 minutes, respectively. Figure 4 shows the Hurst exponent calculated with 6,000 data points by shifting  approximately 500 minutes for the $S\&P$ 500 5-minute index, 
from 1995 to 2004. The average Hurst exponent $H \approx 0.5$ for the $S\&P$ 500 index
sign series of the returns, and $H \approx 0.7$ for the magnitude time series. In addition, the surrogate test shows that the non-linearity of 
the original time series is reflected by the magnitude time series.

Figure 5 shows the Hurst exponent calculated with 6,000 data points by shifting approximately 100 minutes for the KOSDAQ 1-minute index from 1997 to 2004 . 
Though on average $H \approx 0.5$, the Hurst exponent of the returns changes considerably over time, unlike the $S\&P$ 500 index with
a more or less uniform Hurst exponent. In particular, in the KOSDAQ index during its bubble period from the second half 
of 1999 to mid-2000, a large long-term-memory property is observed in the return series. After the market bubble burst, 
we found that the Hurst exponent of the returns dropped to 0.5. This result indicates that the KOSDAQ index may have improved its market 
efficiency after the bubble. As in the previous results, the non-linearity of the original time series of the KOSDAQ data
is reflected in the magnitude time series, and the linearity in the sign time series.

\section{Conclusion}

In this paper, we have investigated the statistical features of international stock market indices of seven countries, high-frequency $S\&P$ 500 and KOSDAQ data. For this purpose, 
the tail index was studied through a linear fitting method by using the Hurst exponent by the 
DFA method. Also, the non-linearity was measured through the surrogate test method. 
We find that the absolute value distribution of the returns of international stock market indices follows a universal power-law distribution, 
having a tail index $\zeta \approx 3$ . However, the Korean stock market follows an exponential distribution with $\beta \approx 0.7$, 
not a power-law distribution.

We also found that in the time series of international market indices, the $S\&P$ 500 index
and the KOSDAQ index, the returns and sign series 
follow random walks ($H \approx 0.5$), but the magnitude series does not. On the other hand, 
we found that in all the time series, the Hurst exponent of the magnitude time series has a long-term-memory property ($0.7 \leq H \leq 0.8$). 
Furthermore, we found that in high-frequency data, the KOSDAQ index, one of the emerging markets, shows higher volatility than a mature market such as the {S\&P} 500 index, which is possibly caused by the abnormally generated bubble. We found a long-term-memory property in the magnitude time series of all data, 
regardless of nation or time scale. Non-linear features of the returns are generally observed in the magnitude time series. 
However, the degree of distribution and correlation in the returns of all data differ in emerging and mature markets. 
Our results may be useful in analyzing global financial markets, for example, differentiating the mature and emerging markets.

\begin{acknowledgements}
This work was supported by a grant from the MOST/KOSEF to the National Core Research Center for Systems Bio-Dynamics (R15-2004-033), and 
by the Ministry of Science \& Technology through the National Research Laboratory Project, and by the Ministry of Education through the program BK 21.
\end{acknowledgements}

\end{document}